Д.Б. Куватова[1,*], Т.П. Панамарев[1], М.В. Ищенко[3,1],
А.П. Глущенко[1], П.П. Берцик[3,1]

[1] В.Г. Фесенков атындағы Астрофизикалық институты, Қазақстан, Алматы қ.
[2] Хериот-Уатт Халықаралық факультеті, Қ. Жұбанов атындағы Ақтөбе өңірлік университеті, Қазақстан, Ақтөбе қ.
[3] Басты астрономиялық обсерватория, Украина Ұлттық ғылым академиясы, Украина, Киев қ.
*e-mail: kuvatova@fai.kz

**Галактикалық орталыққа жақын орналасқан шар тәрізді шоғырлардан шыққан миллисекундтық пульсарлардың гамма-сәулелену ағынын бағалау**

**Аннотация**: Бұл зерттеуде біз миллисекундтық пульсарлардың (МСП) Галактикалық орталықта бақыланатын гамма-сәулеленудің артық мөлшеріне қосқан үлесін қарастырамыз. Ядролық жұлдыздық шоғырмен жақын соқтығысатын алты шар тәрізді шоғырлардың (ШТШ) тікелей N-денелі модельдеуі арқылы алынған жоғары дәлдіктегі симуляциялар деректерін талдаймыз. φ-GPU кодын қолдана отырып, біз симуляциялар барысында қалыптасқан жеке нейтрондық жұлдыздардың (НЖ) орбиталарын барладық, олардың кейбіреулері МСП-ға айналады деп болжаймыз. Осы модельде нейтрондық жұлдыздардың қалыптасу сценарийлерін ескеретін жаңартылған Single Stellar Evolution (SSE) коды қолданылады. Біз бұл МСП-дан гамма-сәулелену ағынын олардың гамма-сәулеленуінің белгілі мәндерін ескере отырып бағаладық. Нәтижелеріміз алты модельденген ШТШ-дан шыққан МСП гамма-сәулелену ағынына аз, бірақ елеулі үлес қосатынын көрсетеді. Осы тұжырыммен бүкіл ШТШ популяциясын ескере отырып, МСП-дан гамма-сәулелену ағыны әлдеқайда жоғары болуы мүмкін екенін және гамма-сәулеленудің артық мөлшеріне ықпал етуі мүмкін екенін болжаймыз. Бұл зерттеу жақын маңдағы шар тәрізді шоғырлардан шыққан МСП-ларды Галактикалық орталықтағы гамма-сәулеленудің артық мөлшерінің әлеуетті көзі ретінде қарастырудың маңыздылығын көрсетеді. Болашақ зерттеулер қос жұлдыздардың эволюциясын ескеретін күрделірек симуляцияларды жүргізуді көздейді және біздің моделімізді нақтылау және бағалауларымыздың дәлдігін арттыру үшін ШТШ-да шынайы бақыланатын МСП үлесін салыстыруды қамтиды.

**Кілт сөздер:** Миллисекундтық пульсарлар, артық мөлшердегі гамма-сәулелену, галактикалық орталық, шар тәрізді шоғырлар, нейтрондық жұлдыздар, N-денелі модельдеу.

**Оценка потока гамма-излучения от миллисекундных пульсаров, происходящих из шаровых скоплений вблизи Галактического центра**

**Аннотация**: В этом исследовании мы рассматриваем вклад миллисекундных пульсаров (МСП) в избыток гамма-излучения, наблюдаемый в Галактическом центре, анализируя данные симуляций высокого разрешения, полученные методом прямого N-тельного моделирования шести шаровых скоплений (ШС), которые испытывают близкие столкновения с ядерным звездным скоплением. Используя код $\varphi$-GPU, мы отследили орбиты отдельных нейтронных звезд (НЗ), образованных в ходе симуляций, предполагая, что часть этих НЗ эволюционирует в МСП. Наша модель включает обновленный код Single Stellar Evolution (SSE), который учитывает сценарии формирования нейтронных звезд. Мы оценили поток гамма-излучения от этих МСП, учитывая известные значения их гамма-излучения. Наши результаты показывают, что МСП, происходящие из шести смоделированных ШС, вносят небольшой, но заметный вклад в наблюдаемый поток гамма-излучения. Этот вывод предполагает, что реальный поток гамма-излучения от МСП может быть намного выше при учете всей популяции ШС, что потенциально может способствовать избытку гамма-излучения. Это исследование подчеркивает важность учета МСП в Галактическом центре, происходящих из близлежащих шаровых скоплений, как потенциального источника наблюдаемого избытка гамма-излучения. Будущие работы будет включать более сложные симуляции, учитывающие эволюцию двойных звезд, и сравнение доли МСП в наблюдаемых ШС для уточнения наших моделей и повышения точности наших оценок.

**Ключевые слова:** Миллисекундные пульсары, избыток гамма-излучения, галактический центр, шаровые скопления, нейтронные звезды, моделирование N-тел.

**D.B. Kuvatova[1,*]** 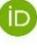 **, T.P. Panamarev[1]** 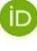 **, M.V. Ishchenko[3,1]** 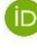 **,**

**A.P. Gluchshenko[1]** 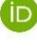 **, P.P. Berczik[3,1]**

[1] Fesenkov Astrophysical Institute, Almaty, Kazakhstan
[2] Heriot-Watt International Faculty, Zhubanov University, Aktobe, Kazakhstan
[3] Main Astronomical Observatory, National Academy of Sciences of Ukraine, Kyiv, Ukraine
*e-mail: kuvatova@fai.kz

## Estimating Gamma-Ray Flux from Millisecond Pulsars Originating in Globular Clusters Near the Galactic Center

**Abstract:** In this study, we investigate the contribution of millisecond pulsars (MSPs) to the gamma-ray excess observed in the Galactic Center by analyzing data from high-resolution direct N-body simulations of six globular clusters (GCs) that experience close encounters with the nuclear star cluster. Using the φ-GPU code, we tracked the orbits of individual neutron stars (NSs) formed during the simulations, assuming a fraction of these NSs evolve into MSPs. Our model includes state-of-the-art single stellar evolution code including prescription for neutron star formation. We estimated the gamma-ray flux from these MSPs, considering known values for their gamma-ray emission. Our results show that MSPs originating from the six modeled GCs contribute a small but non-negligible fraction of the observed gamma-ray flux. This finding suggests that the actual gamma-ray flux from MSPs could be much higher when considering the entire population of GCs, potentially significantly contributing to the gamma-ray excess. This study highlights the importance of considering MSPs in the Galactic Center, originating from nearby globular clusters, as a potential source of the observed gamma-ray excess. Future work will involve more sophisticated simulations incorporating binary stellar evolution and comparing the fraction of MSPs in observed GCs to refine our models and improve the accuracy of our estimates.

**1 Introduction.**
An excess of gamma rays emanating from the central region of the Galaxy (so-called the Galactic Center Excess – GCE) was detected during the analysis of data obtained with the Fermi

Large Area Telescope [1,2]. There are several hypotheses regarding the nature of the GCE, for example, annihilation of dark matter particles [3], radiation from millisecond pulsars [4], and other mechanisms. Such diversity is due to the fact that the issues of the GCE spectrum and spatial morphology [5] are not fully resolved, since different interstellar emission models (IEMs) correspond to completely different characteristics of the GCE [6]. Some authors propose multicomponent models to explain the GCE. For example, the authors [7] consider various IEMs, including the distribution of point sources, interstellar gas, the inverse Compton effect, bremsstrahlung, etc. Their templates with dark matter and millisecond pulsar components showed good agreement with the observed excess. It is likely that several factors are responsible for the GCE, and different components of IEMs contribute to it

The spectral energy distribution of the GCE peaks at several GeV. Since millisecond pulsars have a rapid fall-off above several GeV, they are assumed to be potential sources of the GCE [3]. In addition, it is likely that the spatial distribution of the excess is not spherical and smooth, and correlates with the distribution of stars in the Galactic bulge and bar [8,9]. However, the observed population of millisecond pulsars measured by the Fermi telescope with sufficiently hard spectral indices is insufficient to explain the excess [3,10]. Nevertheless, if pulsars from globular clusters of the Galaxy are taken into account, it might be possible to explain the excess, although current observational data still do not allow making final conclusions.

Globular clusters (GCs) are gravitationally bound star systems, the typical age of which is more than 10-12 billion years [11,12]. Comprehending the relationship between globular clusters and the central areas of galaxies, especially those containing nuclear star clusters and supermassive black holes, is crucial for understanding models of galactic formation [13,14]. Due to dynamical friction, globular clusters tend to migrate to the center of the Galaxy (e.g. [15,16]. Analysis of the GAIA catalogs has shown that the existing globular clusters of the Milky Way feature a wide range of orbits [17,18]. In our previous works, we reconstructed the orbital evolution of observed clusters by backward integration in a time-varying Galactic potential taken from Illustris-TNG cosmological simulations (add links to the 1st papers). We showed that some of the objects approach the Galactic center to within 100 pc [19]. Recently, we conducted a detailed analysis of the clusters that approach the galactic center most closely: we modeled their evolution from 2 Gyr after the formation up to the present day, including stellar evolution [20].

In this paper, we focus on the dynamics of neutron stars (NS) from these globular clusters. By assuming that some of these NSs will result in millisecond pulsars, we analyze gamma-ray emission of these pulsars and discuss their contribution to the observed gamma-ray excess.

The paper is organized as follows. In Section 2 we summarize methods of integration and briefly describe the clusters we studied. In Section 3, we present the main results, and in Section 4 we summarize our findings.

**2 Methods.**

In our previous work [21], we integrated the orbits of 159 GCs, using data from the GC catalog[1] [17,18], transforming them to Galactocentric coordinates [22]. We considered the following parameters: galactocentric distance of the Sun $R_\odot$ = 8.178 kpc [23], height above the galactic plane $Z_\odot$ = 20.8 pc [24] and speed relative to the Local Standard of Rest (LSR) $V_{LSR}$ = 234.737 km/s [25,26].

We assessed the influence of measurement errors on the initial data and GC orbits, finding that the errors had an insignificant impact on many GCs [21].For the integration of GC orbits 10 billion years back in time, we used a dynamically evolving potential from the IllustrisTNG-100 cosmological modeling database [27]. The spatial scales of the disk and dark matter halo were derived using the Miyamoto-Nagai profile [28] for the disk and the Navarro-Frank-White profile [29] for the halo. A detailed procedure for selecting potentials and constructing their components is given in [21,30].

---

[1] https://people.smp.uq.edu.au/HolgerBaumgardt/globular/orbits_table.txt

For numerical simulation, we employed a parallel dynamic N-body code $\varphi$-GPU [31,32], based on a fourth-order Hermite integration scheme with hierarchical individual block time steps. The orbits of the 159 GCs were reconstructed in one of the selected external dynamic potentials, designated as number 411321 in IllustrisTNG-100. From this analysis, we selected six clusters – Palomar 6, HP 1, NGC 6401, NGC 6642, NGC 6681, and NGC 6981 – that have a high probability of approaching the Galactic Center during their dynamic evolution.

We integrated a stellar evolution library into the code [33,34], taking into account stellar winds, supernova core-collapse, and other phenomena. The stellar mass distribution was based on the initial mass function [35] with limits of 0.08-100 $M_\odot$, with each simulated particle corresponding to 10 stars of the same type. This approach, previously used in the works [36,37], allows for more accurate models to determine the necessary parameters for the GCs.

Each GC was initialized to an equilibrium state using the King model distribution function [38], based on such parameters as the half-mass radius $r_{hm}$ and the dimensionless central potential $W_0$. The initial conditions for this simulation were taken from the previous integration without stellar evolution. The Galactic Dynamical Potential was complemented by a Plummer-type potential [39] for the NSC. The supermassive black hole as a separate particle was not included in the integration.

For each of the six clusters, we carried out a series of integrations from 8 billion years ago, varying parameters such as mass, half-mass radius, and King's concentration parameter, to find models that matched the observed values within a deviation of no more than 5%.

**Results and discussion.**

Millisecond pulsars, which are highly energetic remnants from neutron stars, typically form in binary systems with red giants or main-sequence stars (see [40] for a review). Due to the asymmetry of the supernova explosion, newly formed neutron stars experience high kick velocities, which were initially thought to be sufficient for them to escape the low escape velocities of globular clusters. However, observations have confirmed the presence of millisecond pulsars within globular clusters, indicating that they can form and be retained in these environments (see e.g. [41] for a recent discovery).

Our computational code phi-GPU, although not featuring binary stellar evolution, includes state-of-the-art single stellar evolution with scenarios for neutron star formation without any kicks. Using this model, we tracked the orbits of all individual neutron stars formed during the simulations to investigate the dynamics and potential presence of neutron stars that could evolve into millisecond pulsars in the central region of the Milky Way.

We modeled six globular clusters that come close to the Galactic Center: NGC 6642, NGC 6401, HP 1, NGC 6681, Palomar 6, and NGC 6981. Figure 1 shows the positions of all neutron stars from these clusters within a 1 kpc sphere around the Galactic Center, projected in Galactic coordinates with a ±8 degrees box. Neutron stars with blueshifts are shown in blue, while those with redshifts are depicted in red. The figure includes 2210 neutron stars from NGC 6642 (with 1110 still inside the cluster), 380 from NGC 6401, 420 from HP 1, 40 from NGC 6681, and 270 from Palomar 6. For clarity, only every 10th neutron star is displayed. Notably, no neutron stars from NGC 6981 are present in this region. The significant number of neutron stars from NGC 6642 is expected, as the cluster's center frequently lies within this 1 kpc Galactic central region, as indicated by the dense red clump in the plot's upper right.

Thus, Figure 1 demonstrates that a significant number of potential millisecond pulsar progenitors are dispersed within the central kiloparsec of the Galaxy. Assuming that a certain fraction of these neutron stars are actually millisecond pulsars (10%) and using known values for the gamma-ray flux they emit, we can estimate the morphology of this radiation.

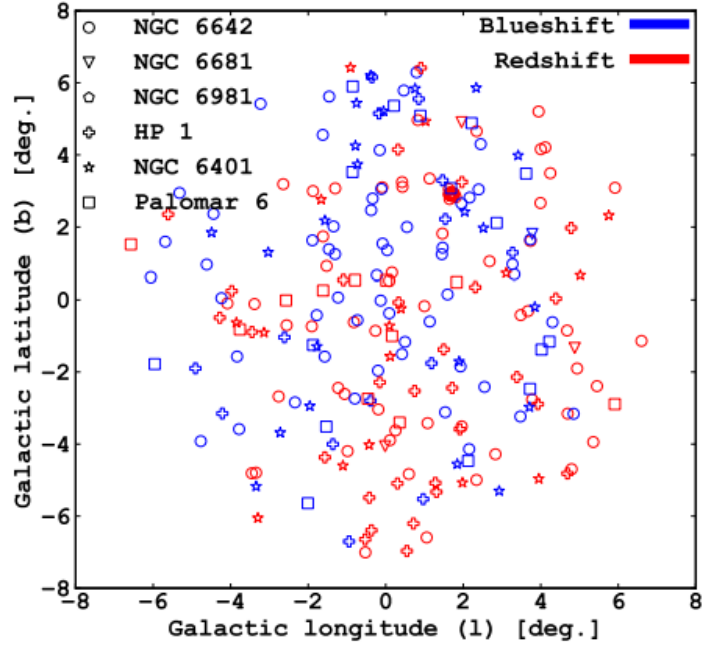

**Figure 1** – Neutron star distributions within a 1 kpc sphere from the GalC are projected in galactic coordinates, with a ±8-degrees box for all GCs at present day. Taken from [20].

Based on the NS distribution obtained from GCs simulations in our simulations, we estimated the gamma-ray flux from millisecond pulsars. We constructed 2D histogram of NSs in the region, and considering the average gamma-ray flux from a millisecond pulsar ranges from $10^{-12}$ to $10^{-11}$ erg s$^{-1}$ cm$^{-2}$ [42], and taking into account that each simulated particle corresponds to 10 physical NS and the fraction of millisecond pulsars from the total number of particles is 0.1, we reconstruct the emission flux from the pulsars. This proportion was selected randomly from the NS distribution. Figure 2 shows the resulting radiation fluxes for cases $10^{-12}$ and $10^{-11}$ erg s$^{-1}$ cm$^{-2}$ in an ±8-degrees box around the central region of the MW (these fluxes may be produced by a MSP with Luminosity of $10^{34}$-$10^{35}$ erg s$^{-1}$ at a distance of 8 kpc).

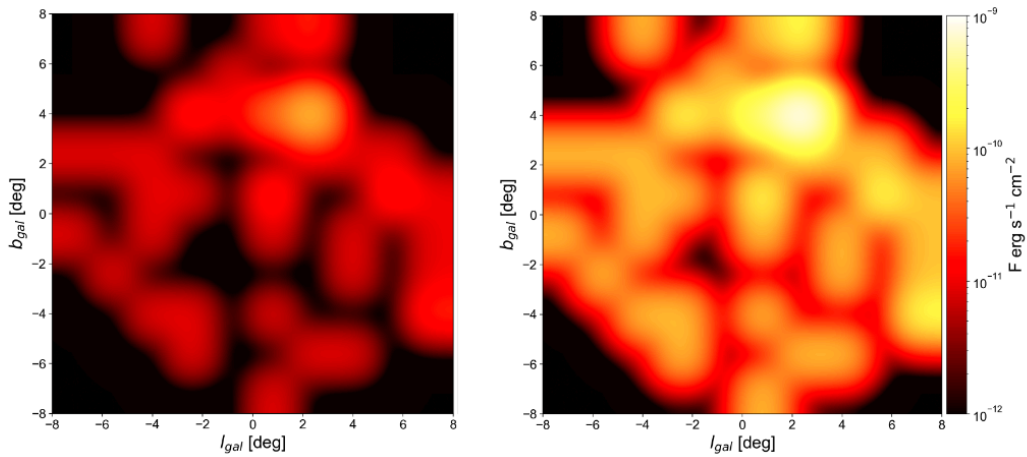

**Figure 2** – Gamma-ray fluxes from the distribution of millisecond pulsars if the average flux from one pulsar is $10^{-12}$ (*left*) and $10^{-11}$ (*right*) in a ±8-degrees box around GalC.

The figure reveals that emission of the MSPs originating from 6 globular clusters contributes to a small fraction of the observed flux [2]. However, we should note that here we made a simple estimation based only on 6 globular clusters that approached the Galactic center in the past that exist in the MW today. However, evidence suggests [43,44] that at least some globular clusters contributed to the formation of the nuclear star cluster of the Milky Way meaning that we see only surviving clusters. Thus, in reality the expectation is that the actual gamma ray flux from MSPs may be much higher, potentially significantly contributing to the observed gamma-ray excess.

**Conclusion.**

We analyzed data from 6 high-resolution direct N-body simulations of Milky Way globular clusters that experience close encounters with the nuclear star cluster [20]. By tracking the orbits of individual neutron stars and assuming a certain fraction of them evolve into millisecond pulsars, we estimated the gamma-ray flux of these pulsars. Although the flux produced by these pulsars is not sufficient to explain the observed gamma-ray excess, our results indicate that the presence of millisecond pulsars in the Galactic Center, originating from nearby globular clusters, should be considered.

To improve the calculations of the contribution of millisecond pulsar flux to the observed gamma-ray radiation, more sophisticated simulations are needed in the future. These simulations should include binary stellar evolution to refine the estimates of the fraction of millisecond pulsars among neutron stars. Additionally, comparing the fraction of millisecond pulsars in observed globular clusters with high numbers of millisecond pulsars can further enhance our models.

**Acknowledgments**. This research has been funded by the Ministry of Science and Higher Education of the Republic of Kazakhstan (Grant No. AP14870501).

**Information about authors:**
**Dana Kuvatova** – MSc; Junior researcher at Fesenkov Astrophysical Institute, Kazakhstan; kuvatova@fai.kz; https://orcid.org/0000-0002-5937-4985;
**Taras Panamarev** – Ph.D; Lead Researcher at Fesenkov Astrophysical Institute, Kazakhstan; Postdoctoral Research Assistant at the University of Oxford, Great Britain; panamarev@aphi.kz, https://orcid.org/0000-0002-1090-4463;
**Maryna Ishchenko** – c.ph.-m.sc. Ph.D; Senior Researcher at Main Astronomical Observatory, National Academy of Sciences of Ukraine; marina@mao.kiev.ua; https://orcid.org/0000-0002-6961-8170;
**Anton Gluchshenko** – MSc-student; Laboratory assistant at Fesenkov Astrophysical Institute, gluchshenko@fai.kz; https://orcid.org/0000-0002-0738-7725;
**Peter Berczik** – Dr.Sci.; Head of department, Main Astronomical Observatory, National Academy of Sciences of Ukraine; berczik@mao.kiev.ua; https://orcid.org/0000-0002-5004-199X.

**References:**

1. Boyarsky A., Malyshev D., Ruchayskiy O. A comment on the emission from the Galactic Center as seen by the Fermi telescope // Phys. Lett. B. Elsevier BV, 2011. Vol. 705, № 3. P. 165–169.
2. Hooper D., Linden T. Origin of the gamma rays from the Galactic Center // Phys. Rev. D. American Physical Society, 2011. Vol. 84, № 12. P. 123005.
3. Hooper D., Goodenough L. Dark matter annihilation in the Galactic Center as seen by the Fermi Gamma Ray Space Telescope // Phys. Lett. B. 2011. Vol. 697, № 5. P. 412–428.
4. Bartels R., Krishnamurthy S., Weniger C. Strong Support for the Millisecond Pulsar Origin of the Galactic Center GeV Excess // Phys. Rev. Lett. 2016. Vol. 116, № 5. P. 051102.
5. Hooper D. The status of the galactic center gamma-ray excess // SciPost Phys. Proc. 2022.
6. Di Mauro M. Investigating the Fermi Large Area Telescope sensitivity of detecting the characteristics of the Galactic center excess // Phys. Rev. D. American Physical Society, 2020. Vol. 102, № 10. P. 103013.
7. Di Mauro M. Characteristics of the Galactic Center excess measured with 11 years of Fermi-LAT data // Phys. Rev. D. American Physical Society, 2021. Vol. 103, № 6. P. 063029.

8. Bartels R. et al. The Fermi-LAT GeV excess as a tracer of stellar mass in the Galactic bulge // Nature Astronomy. Nature Publishing Group, 2018. Vol. 2, № 10. P. 819–828.
9. Macias O. et al. Strong evidence that the galactic bulge is shining in gamma rays // J. Cosmol. Astropart. Phys. IOP Publishing, 2019. Vol. 2019, № 09. P. 042.
10. Abazajian K.N. The consistency of Fermi-LAT observations of the galactic center with a millisecond pulsar population in the central stellar cluster // J. Cosmol. Astropart. Phys. IOP Publishing, 2011. Vol. 2011, № 03. P. 010.
11. Vandenberg D.A. et al. The Ages of 55 Globular Clusters as Determined Using an Improved delta V_TO^HB Method Along with Color-Magnitude Diagram Constraints, and Their Implications for Broader Issues // arXiv [astro-ph.GA]. 2013.
12. Valcin D. et al. Inferring the Age of the Universe with Globular Clusters // arXiv [astro-ph.CO]. 2020.
13. Bland-Hawthorn J., Gerhard O. The Galaxy in Context: Structural, Kinematic, and Integrated Properties // Annu. Rev. Astron. Astrophys. Annual Reviews, 2016. Vol. 54, № Volume 54, 2016. P. 529–596.
14. Neumayer N., Seth A., Böker T. Nuclear star clusters // å. 2020. Vol. 28, № 1. P. 4.
15. Tremaine S.D. The effect of dynamical friction on the orbits of the Magellanic clouds // Astrophysical Journal, vol. 203, Jan. 1, 1976, pt. adsabs.harvard.edu, 1976.
16. Capuzzo-Dolcetta R. The Evolution of the Globular Cluster System in a Triaxial Galaxy: Can a Galactic Nucleus Form by Globular Cluster Capture? // \apj. 1993. Vol. 415. P. 616.
17. Vasiliev E., Baumgardt H. Gaia EDR3 view on galactic globular clusters // Mon. Not. R. Astron. Soc. Oxford Academic, 2021. Vol. 505, № 4. P. 5978–6002.
18. Baumgardt H., Vasiliev E. Accurate distances to Galactic globular clusters through a combination of Gaia EDR3, HST, and literature data // \mnras. 2021. Vol. 505, № 4. P. 5957–5977.
19. Ishchenko M. et al. Milky Way globular clusters on cosmological timescales - II. Interaction with the Galactic centre // Astron. Astrophys. Suppl. Ser. EDP Sciences, 2023. Vol. 674. P. A70.
20. Ishchenko M. et al. Dynamical evolution of Milky Way globular clusters on the cosmological timescale I. Mass loss and interaction with the nuclear star cluster // arXiv [astro-ph.GA]. 2024.
21. Ishchenko M. et al. Milky Way globular clusters on cosmological timescales - I. Evolution of the orbital parameters in time-varying potentials // Astron. Astrophys. Suppl. Ser. EDP Sciences, 2023. Vol. 673. P. A152.
22. Chemerynska I.V. et al. Kinematic characteristics of the Milky Way globular clusters based on Gaia DR2 data // arXiv [astro-ph.GA]. 2022.
23. Reid M.J., Brunthaler A. The proper motion of Sagittarius A*. ii. The mass of Sagittarius A // Astrophys. J. American Astronomical Society, 2004. Vol. 616, № 2. P. 872–884.
24. Bennett M., Bovy J. Vertical waves in the solar neighbourhood in Gaia DR2 // Mon. Not. R. Astron. Soc. Oxford Academic, 2018. Vol. 482, № 1. P. 1417–1425.
25. Bovy J. et al. The Milky Way's circular-velocity curve between 4 and 14 kpc from APOGEE data // Astrophys. J. IOP Publishing, 2012. Vol. 759, № 2. P. 131.
26. Drimmel R., Poggio E. On the solar velocity // Res. Notes AAS. American Astronomical Society, 2018. Vol. 2, № 4. P. 210.
27. Nelson D. et al. The IllustrisTNG simulations: public data release // Computational Astrophysics and Cosmology. 2019. Vol. 6, № 1. P. 2.
28. Miyamoto M., Nagai R. Three-dimensional models for the distribution of mass in galaxies // , vol. 27, no. 4, 1975, p. 533-543. adsabs.harvard.edu, 1975.
29. Navarro J.F., Frenk C.S., White S.D.M. A universal density profile from hierarchical clustering // Astrophys. J. American Astronomical Society, 1997. Vol. 490, № 2. P. 493–508.
30. Mardini M.K. et al. Cosmological insights into the early accretion of r-process-enhanced stars. I. a comprehensive chemodynamical analysis of LAMOST J1109+0754 // Astrophys. J. American Astronomical Society, 2020. Vol. 903, № 2. P. 88.

31. Berczik P. et al. High performance massively parallel direct N-body simulations on large GPU clusters // International conference on high performance computing. 2011. P. 8–18.
32. Berczik P. et al. Up to 700k GPU Cores, Kepler, and the Exascale Future for Simulations of Star Clusters Around Black Holes // Supercomputing. Springer Berlin Heidelberg, 2013. P. 13–25.
33. Kamlah A.W.H. et al. Preparing the next gravitational million-body simulations: evolution of single and binary stars in NBODY6++GPU, MOCCA, and MCLUSTER // \mnras. 2022. Vol. 511, № 3. P. 4060–4089.
34. Banerjee S. et al. BSE versus StarTrack: Implementations of new wind, remnant-formation, and natal-kick schemes in NBODY7 and their astrophysical consequences // Astron. Astrophys. Suppl. Ser. EDP Sciences, 2020. Vol. 639. P. A41.
35. Kroupa P. On the variation of the initial mass function // Mon. Not. R. Astron. Soc. 2001. Vol. 322, № 2. P. 231–246.
36. Panamarev T. et al. Star-disc interaction in galactic nuclei: formation of a central stellar disc // \mnras. 2018. Vol. 476, № 3. P. 4224–4233.
37. Panamarev T. et al. Direct N-body simulation of the Galactic centre // \mnras. 2019. Vol. 484, № 3. P. 3279–3290.
38. King I.R. The structure of star clusters. III. Some simple dynamical models // AJS. 1966. Vol. 71. P. 64.
39. Plummer H.C. On the problem of distribution in globular star clusters // \mnras. 1911. Vol. 71. P. 460–470.
40. Manchester R.N. Millisecond Pulsars, their Evolution and Applications // J. Astrophys. Astron. 2017. Vol. 38, № 3. P. 42.
41. Padmanabh P.V. et al. Discovery and timing of ten new millisecond pulsars in the globular cluster Terzan 5 // Astron. Astrophys. Suppl. Ser. EDP Sciences, 2024. Vol. 686. P. A166.
42. Xing Y., Wang Z. FERMI STUDY OF γ-RAY MILLISECOND PULSARS: THE SPECTRAL SHAPE AND PULSED EMISSION FROM J0614–3329 UP TO 60 GeV // ApJ. IOP Publishing, 2016. Vol. 831, № 2. P. 143.
43. Do T. et al. Revealing the Formation of the Milky Way Nuclear Star Cluster via Chemo-dynamical Modeling // \apjl. 2020. Vol. 901, № 2. P. L28.
44. Arca Sedda M. et al. On the Origin of a Rotating Metal-poor Stellar Population in the Milky Way Nuclear Cluster // \apjl. 2020. Vol. 901, № 2. P. L29.

**References:**

1. Boyarsky, A., Malyshev, D. & Ruchayskiy, O. A comment on the emission from the Galactic Center as seen by the Fermi telescope. *Phys. Lett. B* **705**, 165–169 (2011).
2. Hooper, D. & Linden, T. Origin of the gamma rays from the Galactic Center. *Phys. Rev. D* **84**, 123005 (2011).
3. Hooper, D. & Goodenough, L. Dark matter annihilation in the Galactic Center as seen by the Fermi Gamma Ray Space Telescope. *Phys. Lett. B* **697**, 412–428 (2011).
4. Bartels, R., Krishnamurthy, S. & Weniger, C. Strong Support for the Millisecond Pulsar Origin of the Galactic Center GeV Excess. *Phys. Rev. Lett.* **116**, 051102 (2016).
5. Hooper, D. The status of the galactic center gamma-ray excess. *SciPost Phys. Proc.* (2022) doi:10.21468/scipostphysproc.12.006.
6. Di Mauro, M. Investigating the Fermi Large Area Telescope sensitivity of detecting the characteristics of the Galactic center excess. *Phys. Rev. D* **102**, 103013 (2020).
7. Di Mauro, M. Characteristics of the Galactic Center excess measured with 11 years of Fermi-LAT data. *Phys. Rev. D* **103**, 063029 (2021).
8. Bartels, R., Storm, E., Weniger, C. & Calore, F. The Fermi-LAT GeV excess as a tracer of stellar mass in the Galactic bulge. *Nature Astronomy* **2**, 819–828 (2018).
9. Macias, O. *et al.* Strong evidence that the galactic bulge is shining in gamma rays. *J. Cosmol.*

*Astropart. Phys.* **2019**, 042 (2019).
10. Abazajian, K. N. The consistency of Fermi-LAT observations of the galactic center with a millisecond pulsar population in the central stellar cluster. *J. Cosmol. Astropart. Phys.* **2011**, 010 (2011).
11. Vandenberg, D. A., Brogaard, K., Leaman, R. & Casagrande, L. The Ages of 55 Globular Clusters as Determined Using an Improved delta V_TO^HB Method Along with Color-Magnitude Diagram Constraints, and Their Implications for Broader Issues. *arXiv [astro-ph.GA]* (2013) doi:10.1088/0004-637X/775/2/134.
12. Valcin, D., Bernal, J. L., Jimenez, R., Verde, L. & Wandelt, B. D. Inferring the Age of the Universe with Globular Clusters. *arXiv [astro-ph.CO]* (2020) doi:10.1088/1475-7516/2020/12/002.
13. Bland-Hawthorn, J. & Gerhard, O. The Galaxy in Context: Structural, Kinematic, and Integrated Properties. *Annu. Rev. Astron. Astrophys.* **54**, 529–596 (2016).
14. Neumayer, N., Seth, A. & Böker, T. Nuclear star clusters. *å* **28**, 4 (2020).
15. Tremaine, S. D. The effect of dynamical friction on the orbits of the Magellanic clouds. *Astrophysical Journal, vol. 203, Jan. 1, 1976, pt* (1976).
16. Capuzzo-Dolcetta, R. The Evolution of the Globular Cluster System in a Triaxial Galaxy: Can a Galactic Nucleus Form by Globular Cluster Capture? *\apj* **415**, 616 (1993).
17. Vasiliev, E. & Baumgardt, H. Gaia EDR3 view on galactic globular clusters. *Mon. Not. R. Astron. Soc.* **505**, 5978–6002 (2021).
18. Baumgardt, H. & Vasiliev, E. Accurate distances to Galactic globular clusters through a combination of Gaia EDR3, HST, and literature data. *\mnras* **505**, 5957–5977 (2021).
19. Ishchenko, M., Sobolenko, M., Kuvatova, D., Panamarev, T. & Berczik, P. Milky Way globular clusters on cosmological timescales - II. Interaction with the Galactic centre. *Astron. Astrophys. Suppl. Ser.* **674**, A70 (2023).
20. Ishchenko, M. *et al.* Dynamical evolution of Milky Way globular clusters on the cosmological timescale I. Mass loss and interaction with the nuclear star cluster. *arXiv [astro-ph.GA]* (2024).
21. Ishchenko, M. *et al.* Milky Way globular clusters on cosmological timescales - I. Evolution of the orbital parameters in time-varying potentials. *Astron. Astrophys. Suppl. Ser.* **673**, A152 (2023).
22. Chemerynska, I. V., Ishchenko, M. V., Sobolenko, M. O., Khoperskov, S. A. & Berczik, P. P. Kinematic characteristics of the Milky Way globular clusters based on Gaia DR2 data. *arXiv [astro-ph.GA]* (2022).
23. Reid, M. J. & Brunthaler, A. The proper motion of Sagittarius A*. ii. The mass of Sagittarius A. *Astrophys. J.* **616**, 872–884 (2004).
24. Bennett, M. & Bovy, J. Vertical waves in the solar neighbourhood in Gaia DR2. *Mon. Not. R. Astron. Soc.* **482**, 1417–1425 (2018).
25. Bovy, J. *et al.* The Milky Way's circular-velocity curve between 4 and 14 kpc from APOGEE data. *Astrophys. J.* **759**, 131 (2012).
26. Drimmel, R. & Poggio, E. On the solar velocity. *Res. Notes AAS* **2**, 210 (2018).
27. Nelson, D. *et al.* The IllustrisTNG simulations: public data release. *Computational Astrophysics and Cosmology* **6**, 2 (2019).
28. Miyamoto, M. & Nagai, R. Three-dimensional models for the distribution of mass in galaxies. *, vol. 27, no. 4, 1975, p. 533-543.* (1975).
29. Navarro, J. F., Frenk, C. S. & White, S. D. M. A universal density profile from hierarchical clustering. *Astrophys. J.* **490**, 493–508 (1997).
30. Mardini, M. K. *et al.* Cosmological insights into the early accretion of r-process-enhanced stars. I. a comprehensive chemodynamical analysis of LAMOST J1109+0754. *Astrophys. J.* **903**, 88 (2020).
31. Berczik, P. *et al.* High performance massively parallel direct N-body simulations on large GPU clusters. in *International conference on high performance computing* 8–18 (2011).
32. Berczik, P. *et al.* Up to 700k GPU Cores, Kepler, and the Exascale Future for Simulations of

Star Clusters Around Black Holes. in *Supercomputing* 13–25 (Springer Berlin Heidelberg, 2013). doi:10.1007/978-3-642-38750-0_2.
33. Kamlah, A. W. H. *et al.* Preparing the next gravitational million-body simulations: evolution of single and binary stars in NBODY6++GPU, MOCCA, and MCLUSTER. \mnras **511**, 4060–4089 (2022).
34. Banerjee, S. *et al.* BSE versus StarTrack: Implementations of new wind, remnant-formation, and natal-kick schemes in NBODY7 and their astrophysical consequences. *Astron. Astrophys. Suppl. Ser.* **639**, A41 (2020).
35. Kroupa, P. On the variation of the initial mass function. *Mon. Not. R. Astron. Soc.* **322**, 231–246 (2001).
36. Panamarev, T. *et al.* Star-disc interaction in galactic nuclei: formation of a central stellar disc. \mnras **476**, 4224–4233 (2018).
37. Panamarev, T. *et al.* Direct N-body simulation of the Galactic centre. \mnras **484**, 3279–3290 (2019).
38. King, I. R. The structure of star clusters. III. Some simple dynamical models. *AJS* **71**, 64 (1966).
39. Plummer, H. C. On the problem of distribution in globular star clusters. \mnras **71**, 460–470 (1911).
40. Manchester, R. N. Millisecond Pulsars, their Evolution and Applications. *J. Astrophys. Astron.* **38**, 42 (2017).
41. Padmanabh, P. V. *et al.* Discovery and timing of ten new millisecond pulsars in the globular cluster Terzan 5. *Astron. Astrophys. Suppl. Ser.* **686**, A166 (2024).
42. Xing, Y. & Wang, Z. FERMI STUDY OF γ-RAY MILLISECOND PULSARS: THE SPECTRAL SHAPE AND PULSED EMISSION FROM J0614–3329 UP TO 60 GeV. *ApJ* **831**, 143 (2016).
43. Do, T. *et al.* Revealing the Formation of the Milky Way Nuclear Star Cluster via Chemo-dynamical Modeling. \apjl **901**, L28 (2020).
44. Arca Sedda, M. *et al.* On the Origin of a Rotating Metal-poor Stellar Population in the Milky Way Nuclear Cluster. \apjl **901**, L29 (2020).